\documentclass[10pt, two column, twoside]{IEEEtran}
\usepackage{color}
\usepackage[colorlinks, linkcolor=color1, anchorcolor=blue, citecolor=color1]{hyperref}

\usepackage{amssymb}
\usepackage{amsmath}
\usepackage[lined,boxed,commentsnumbered, ruled]{algorithm2e}
\usepackage{mathrsfs}
\usepackage{algorithmic}
\usepackage{bm}
\usepackage{hhline}

\usepackage{pgfplots}
\usepackage{tikz}
\usetikzlibrary{arrows}
\usepackage{subfigure}
\usepackage{graphicx,booktabs,multirow}

\definecolor{colorhkust}{RGB}{20,43,140}
\definecolor{colortsinghua}{RGB}{116,52,129}
\definecolor{color1}{RGB}{128,0,0}

\date{}

\begin{document}

        \title{Integrated Sensing-Communication-Computation for Edge Artificial Intelligence}
\author{Dingzhu Wen, Xiaoyang Li, Yong Zhou, Yuanming Shi, Sheng Wu, and Chunxiao Jiang
        \thanks{D. Wen, Y. Zhou, and Y. Shi are with School of Information Science and Technology, ShanghaiTech University, Shanghai 201210, China (e-mail: \{wendzh,zhouyong,shiym\}@shanghaitech.edu.cn).}
		\thanks{X. Li is with Shenzhen Research Institute of Big Data, Shenzhen 518172, China (e-mail:lixiaoyang@sribd.cn).}
		\thanks{S. Wu is with the School of Information and Communication Engineering, Beijing University of Posts and Telecommunications, Beijing 100876, China (e-mail: thuraya@bupt.edu.cn).}
        \thanks{C. Jiang is with the Tsinghua Space Center and the Beijing National Re- search Center for Information Science and Technology, Tsinghua University, Beijing 100084, China (e-mail: jchx@tsinghua.edu.cn).}
                }

\maketitle


\begin{abstract}
\textcolor{blue}{Edge artificial intelligence (AI) has been a promising solution towards 6G to empower a series of advanced techniques such as digital twins, holographic projection, semantic communications, and auto-driving, for achieving intelligence of everything.} The performance of edge AI tasks, including edge learning and edge AI inference, depends on the quality of three highly coupled processes, i.e., sensing for data acquisition, computation for information extraction, and communication for information transmission. However, these three modules need to compete for network resources for enhancing their own quality-of-services. To this end, integrated sensing-communication-computation (ISCC) is of paramount significance for improving resource utilization as well as achieving the customized goals of edge AI tasks. By investigating the interplay among the three modules, this article presents various kinds of ISCC schemes for federated edge learning tasks and edge AI inference tasks in both application and physical layers.
\end{abstract}



\section{Introduction}

\begin{figure}[h]
\centering
\includegraphics[width=0.45\textwidth]{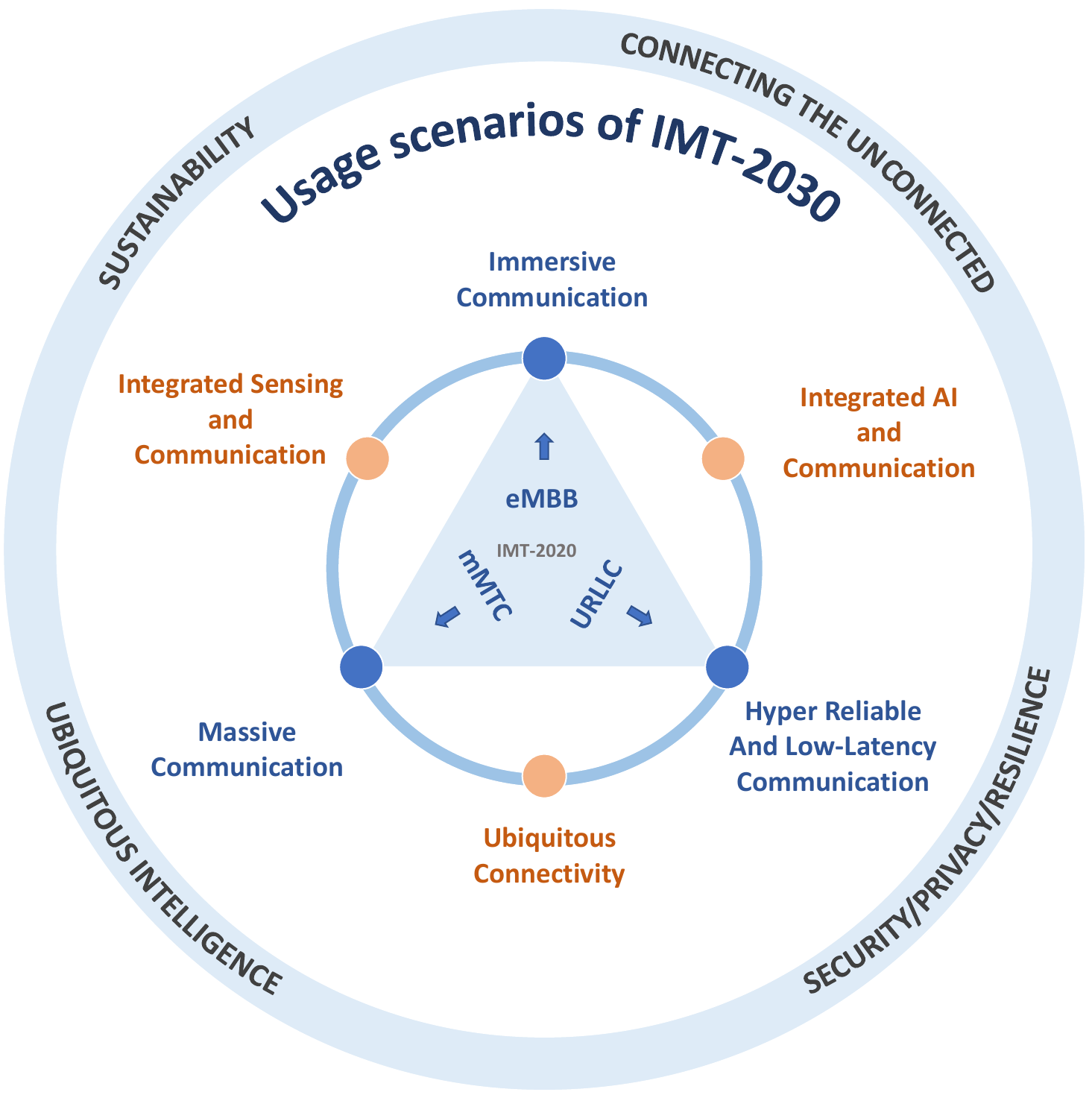}
\caption{Six Usage Scenarios of 6G Proposed by IMT-2030 \cite{IMT2030}.}
\label{6GUsageScenarios}
\end{figure}

By enabling various network architectures like space-air-ground integrated networks, computing power networks, and internet-of-things networks, as well as via leveraging massive distributed computing powers and data resources therein, 6G will go far beyond the traditional data-centric services to push forward intelligence of everything for providing ubiquitous and immersive intelligent services \cite{shi2023task}. {\color{blue} The integration of communication and AI has been recognized as a key usage scenario in 6G by IMT-2030, as shown in Fig. \ref{6GUsageScenarios} \cite{IMT2030}.} This calls for the deployment of artificial intelligence (AI) models at the network edge via edge training and edge inference \cite{letaief2021edge}. Specifically, the collaborative edge training framework called federated edge learning (FEEL) has been a promising solution for efficiently distilling AI models at the network edge with the benefits of fast mobile data access and data privacy preservation. The edge-device collaborative inference technique splits well-trained AI models between edge servers and devices and utilizes their inference capability for providing secure, private, and real-time intelligent services. 

As massive sensors (e.g., radars, cameras, lidars) are deployed at the network edge, the prosperity of edge AI has been promoted by the generated huge amount of sensory data \cite{feng2021joint}. Particularly, the implementation of edge AI tasks requires the fusion of physical, biological, and cyber worlds. They involve three key processes, i.e., sensing for obtaining information from the environment, communication for information sharing, and computation for further processing the information for making intelligent decisions \cite{zhu2023pushing}. These three modules are highly coupled since the task performance depends on all modules’ quality-of-service (QoS) and they should compete for network resources for suppressing their own distortion \cite{wen2022task}. {\color{blue}Existing ISCC schemes ignore the coupling property of the three modules, thus leading to suboptimal performance.} Furthermore, these sensing-based edge AI tasks compete with existing communication and computation services like mobile edge computing for network resources, such as bandwidth, energy, and time \cite{qi2020integrated}. To this end, it calls for integrated sensing-communication-computation (ISCC) schemes to coordinate the three processes in specific edge AI tasks for enhancing their performance as well as to balance the co-existence of the sensing-based edge AI tasks and the conventional tasks for guaranteeing the QoS of all tasks.

The ISCC schemes enjoy the benefits of better network resource coordination among the three modules and hardware sharing between sensing and communication on devices for saving their physical spaces \cite{cui2021integrating}. However, the achievement of these advantages faces new challenges. In the task level, edge AI features a task-oriented property that concerns the effectiveness and efficiency instead of the traditional design criteria such as system throughput and signal-to-noise ratio (SNR). Therefore, they call for new design criteria. As well, their design goals are customized and are lack of a unified form, e.g., the training latency for federated learning tasks and instantaneous inference accuracy for inference tasks. Furthermore, the design complexity of ISCC schemes is much higher than that of the separated schemes due to the joint design of the highly coupled sensing, communication, and computation modules in edge AI tasks. In the physical layer, the co-existence of edge AI tasks and traditional services causes interference between each other. Besides, reusing one waveform to simultaneously perform the functions of sensing, communication, and computation, has been a promising solution for further saving the radio resources. However, this requires high complexity waveform designs for jointly optimizing multiple objects.

\begin{figure*}[t] \centering
	\includegraphics[width=0.8\textwidth]{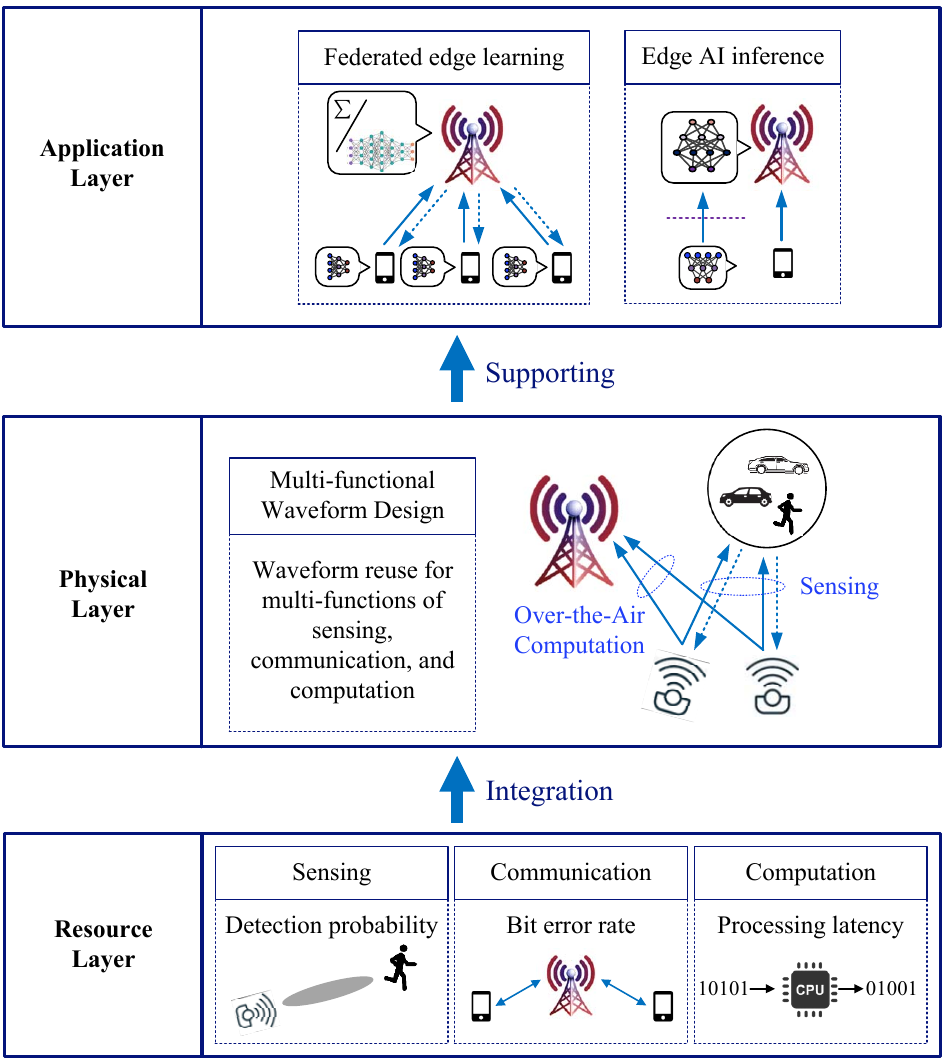}
	\caption{Integrated Sensing-Computation-Communication for Edge AI.}
	\label{Fig:Overview}
\end{figure*}

To address these challenges, this article advocates an ISCC approach for edge AI tasks. We discuss the vision, motivations, and principles of ISCC designs from the task-level resource management and the physical-layer waveform design for FEEL and edge AI inference, as shown in Fig. \ref{Fig:Overview}. Particularly, depending on the underlying transceiver technique, we elaborate the holistic design of both digital and analog ISSC for FEEL tasks that account for limited radio and computation resources. To provide high quality inference services, we investigate the coupling mechanism among sensing, communication, and computation therein, followed by introducing two ISCC schemes for narrow-view sensing based and wide-view sensing based multi-device AI inference tasks, respectively. To fully exploit the radio resource for edge AI, physical layer ISCC techniques are investigated, where the beamformers for multi-functional signals are designed.

\section{ISCC for Federated Edge Learning Tasks}

\begin{figure*}[t] \centering
	\includegraphics[width=1\textwidth]{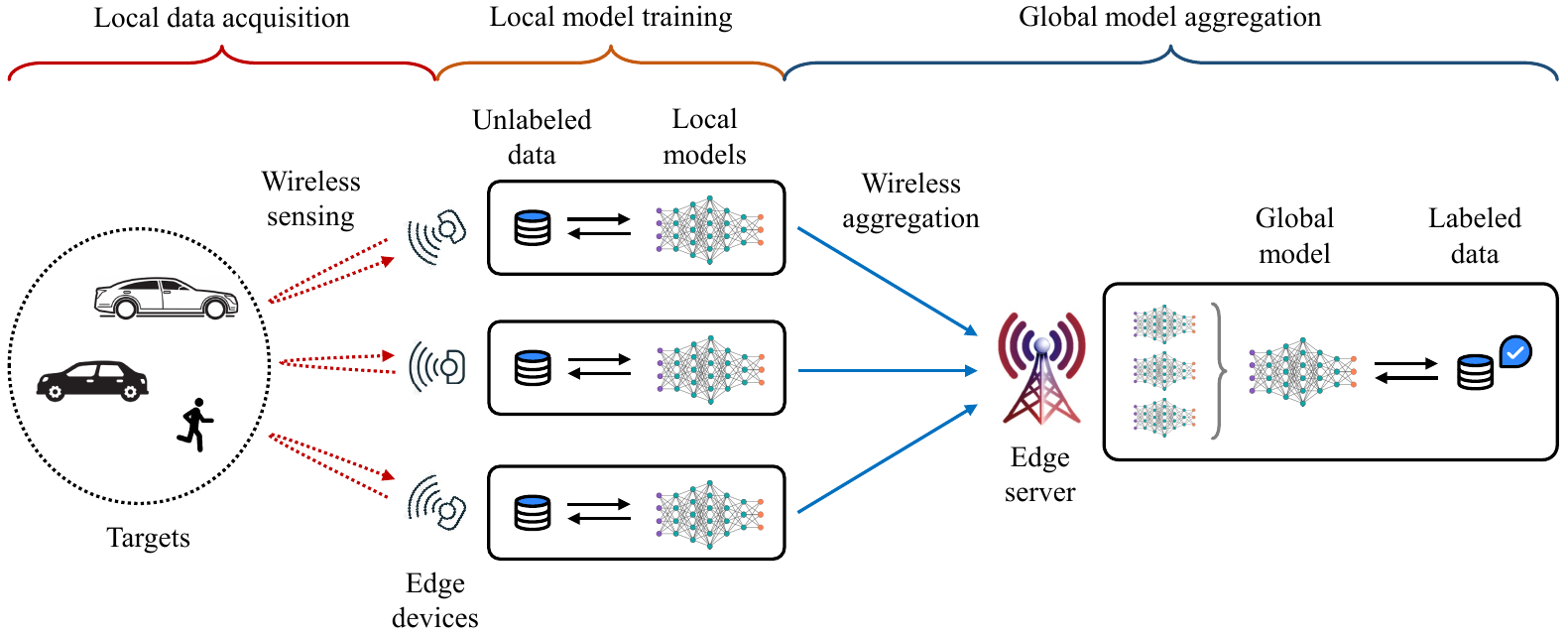}
	\caption{Integrated Sensing-Computation-Communication for Federated Edge Learning.}
	\label{Fig:ISCC_FEEL}
\end{figure*}

\subsection{Vision}
As arguably the most popular distributed ML paradigm, FEEL is featured by the collaborative yet privacy-preserving training of a shared statistical model among multiple geographically dispersed edge devices via only exchanging model updates rather than private data. 
The training of FEEL typically involves the iteration of local model training and global model aggregation. 
During local model training, each of the participating devices updates its local model via minimizing the local loss based on the local dataset and the up-to-date global model, where different algorithms (e.g., zeroth/first/second-order algorithms) can be applied. 
During global model aggregation, all participating devices upload their local models to the edge server, which then aggregates these local models to update the global model that will be disseminated to the devices. 
{\color{blue}Because of the privacy-preserving property, FEEL has found its application in smart healthcare, autonomous driving, smart finance, etc.}


It has been demonstrated that the achievable performance of FEEL (e.g., test accuracy and convergence rate) critically depends on the quality and quantity of the training data available at edge devices, the spectrum resources, and the network topology of the underlying communication systems, as well as the computation resources available at the edge server and devices. 
Most of the existing studies on FEEL focuses on developing effective device scheduling, transceiver design, and knowledge distillation schemes to enhance the learning performance from the communication and computation perspectives, while assuming that the high-quality training data is readily available at each edge device. 
However, this assumption may not hold in many practical scenarios, where the data acquisition and model training processes occur simultaneously. 
In particular, to support ambient intelligence (e.g., extended human sensing and health monitoring) under the FEEL framework, the real-time data acquired by wireless sensing should be utilized for model training to learn responsive and up-to-date models, as shown in Fig. \ref{Fig:ISCC_FEEL}. 
{\color{blue}Specifically, during each training round of FEEL, a certain volume of data are first obtained via wireless sensing to capture the latest status of the sensing targets, and then utilized to train model parameters for identifying essential system characteristics. 
Meanwhile, considering holistic ISCC design is necessary for achieving efficient FEEL due to the coupling of resource consumption among data acquisition, local training, and model transmission.}
Depending on the underlying communication technique, the digital and analog ISCC for FEEL tasks are elaborated in the following two sub-sections. 

%

\subsection{Digital ISCC for Federated Edge Learning}
For FEEL with digital ISCC, the sensing, communication, and computation processes are performed in a sequential manner. 
Specifically, the training process of each iteration consists of three steps, i.e., local data acquisition via wireless sensing, local model training with sensed data, and global model aggregation. 
As a key step that distinguishes from the conventional FEEL, the local data acquisition is generally task-specific. 
Therefore, we take FEEL with digital ISCC for the application of human motion recognition as an example.
Specifically, each device first transmits frequency-modulated continuous wave (FMCW) that consists of multiple up-chirps to the sensing target and then receives the reflected radar echo signal. 
After processing the echo signal with a singular value decomposition (SVD) filter and short time Fourier transformation (STFT), the sensing data is then transformed to the form of spectrogram, facilitating further data utilization \cite{liu2022toward}. 
Without any prior information about the sensing target, the sensing data may not be well-labeled, which is a critical issue that should be considered. 
With unlabeled data, either unsupervised or semi-supervised learning needs to be considered during the local model training step. 
After local model training, the edge devices transmit their up-to-date local models to the edge server using digital modulation, e.g., quadrature amplitude modulation. 
After successfully receiving  all the local models, the edge server performs global model aggregation, where the statistical heterogeneity and system heterogeneity can be alleviated by appropriately designing the aggregation coefficients.


To facilitate the holistic design of digital ISCC for FEEL, it is essential to characterize the impact of the sensing, communication, and computation modules on the ultimate performance of FL tasks, followed by integrating these modules organically and developing a task-oriented resource allocation scheme to enhance the overall learning performance under the communication and computation resource constraints.
{\color{blue} Although some metrics (e.g., structural similarity \cite{liu2022toward}, mean squared error (MSE)) have been proposed for measuring sensing data accuracy, they are unable to characterize the relationship between the wireless sensing signals and the accuracy of data acquisition, which makes it necessary to develop an effective performance metric to quantify the impact of resource allocation and signal design on wireless sensing.} 
In addition, the separate design of sensing, communication, and computation modules may lead to objective inconsistency with the ultimate goal and in turn reduce the overall learning performance, which necessities the joint ISCC design from various perspectives (e.g., waveform design, resource allocation). 
In practical scenarios, the devices located at different locations may sense different types of data, which necessitates ISCC-enabled multi-task FEEL. 
Due to the mobility of sensing targets, a device may not always be able to sense the required data for local training, which makes device scheduling important for managing the sensing tasks to improve the efficiency of resource utilization.

\subsection{Analog ISCC for Federated Edge Learning}
Over-the-air computation (AirComp), as an emerging analog communication scheme, has been widely adopted to enable spectral-efficient and low-latency FEEL (see e.g., \cite{yang2020federated,wang2021federated}. 
Meanwhile, wireless sensing data can be extracted from the analog signals reflected from the sensing targets, whose feature is mainly represented by the signal variation during the propagation process. 
These observations make it possible to exploit analog modulation to simultaneously accomplish the tasks of sensing data acquisition and global model aggregation, where leveraging the shared hardware in each device to enable integrated sensing and communication is essential. 
In this way, each edge device can transmit the current local model to the edge server while sensing new training data simultaneously, thereby having the potential to significantly improve the training efficiency of ISCC-enabled FEEL. 
Although superimposed signals can be naturally aggregated over the air for improving the communication efficiency, the signals reflected from different sensing targets may interfere with each other and then degrade the quality of sensed data. 
Hence, how to implement appropriate transceiver design to align signals for efficient model aggregation while suppressing the interference reflected from different targets and clutters is a challenging issue to be tackled in analog ISCC-enabled FEEL.

Due to the parallel execution of data sensing and model aggregation, it is critical to synchronize the sensing and communication processes. Specifically, the delay of analog model aggregation with AirComp is mainly determined by the dimension of model parameters to be uploaded, while the delay of sensing process is determined by the amount of required sensing data and duration of each sensing slot. To ensure sensing and communication to complete their own tasks at the same time within a given phase time, how to regulate the above factors is also important for analog ISCC-enabled FEEL in addition to the transceiver design. For instance, if the delay of model aggregation is larger than that of data sensing, it is necessary to compress the model parameters to reduce the communication latency, or increase the number of sensing data and sensing duration to increase the sensing time for synchronization. Besides, if the delay of model aggregation is less than that of data sensing in a complicated sensing task, it is important to increase the aggregation time while improving the aggregation accuracy through signal retransmission.
Moreover, to enhance the performance of ISCC-enabled FEEL in mobile scenarios with Doppler frequency shift, orthogonal time frequency space (OTFS), as an emerging modulation technology, can be further integrated with AirComp to simultaneously achieve reliable data sensing and low-latency model aggregation.

\begin{figure*}[t] 
\centering
	\includegraphics[width=1\textwidth]{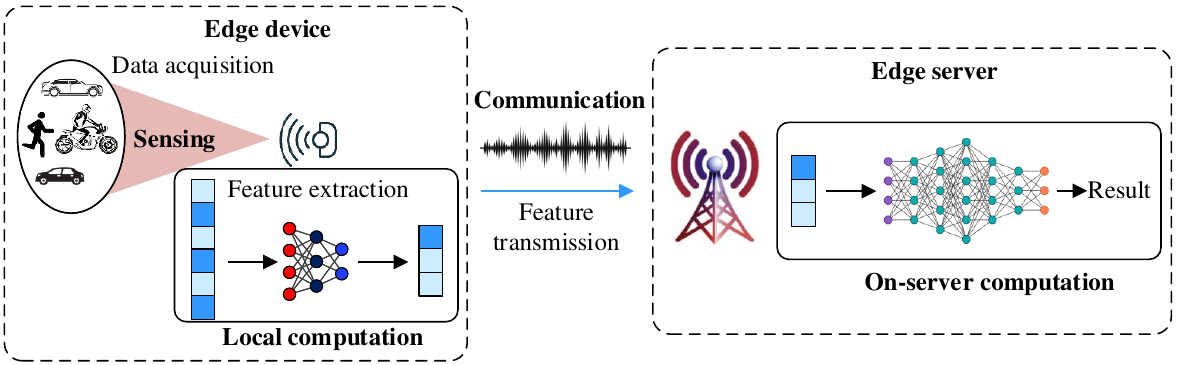}
	\caption{Integrated Sensing-Computation-Communication for Edge-device Collaborative Inference.}
	\label{Fig:Inference}
\end{figure*}

\section{ISCC for Edge AI Inference Tasks}

\subsection{Vision}

{\color{blue}Edge AI inference has been a key technique to deploy well-trained AI models at the network edge to facilitate the fast access of massive mobile data for making real-time intelligent decisions, thereby enabling ubiquitous intelligent services like smart home, eHealth, and auto-driving, with low latency.} Different from the conventional paradigms of on-device inference which locally executes AI models at devices and suffers from high computation overhead, and on-server inference which uploads raw data from edge devices to an edge server for completing inference tasks and violates the data privacy of mobile users, the edge-device collaborative inference paradigm has been a promising solution. It divides an AI model into two parts. The front-end part with light size is deployed at devices for local feature extraction. The back-end part, which requires intensive computation and is deployed at the edge server, receives the local feature vector for completing the downstream inference task. As a result, it enjoys the benefits of data privacy preservation, computation offloading, and scalability in terms of participating devices.

The implementation of edge-device collaborative inference requires the cooperation of three modules at edge devices, i.e., sensing for data acquisition, computation for feature extraction and processing, and communication for feature transmission to the edge server, as shown in Fig. \ref{Fig:Inference}. The inference accuracy is decided by the server-side received feature vector, whose distortion level depends on sensing noise, limited processing capabilities, and hostile wireless channels. Besides, in order to suppress each module's own distortion, sensing, computation, and communication need to compete for the limited network resources like time and energy. What’s more, the dimension of sensory data decides the feature extraction complexity and the size of the extracted feature vector determines the communication load. Hence, sensing, computation, and communication are highly coupled in an edge-device collaborative inference task. It calls for ISCC designs from a systematic view. In the sequel, the ISCC based edge inference framework is first introduced, followed by the elaboration of ISCC schemes for multi-device AI systems.

\subsection{ISCC based Edge Inference Framework}
\subsubsection{Task-oriented Design Criteria}
{\color{blue}In edge inference tasks such as human motion recognition, the conventional design criteria for each module such as sensing’s Cramér–Rao Bounds, computation latency, and communication throughput, do not work well.} This is because they fail to distinguish the feature elements that have same sensing distortion level, processing time, and size but have heterogeneous contributions on the inference accuracy. In other words, edge inference concerns the effective and efficient execution of the well-trained AI models. Hence, it’s desirable to directly adopt inference accuracy as the design goal \cite{shi2023task}. However, the value of the instantaneous inference accuracy is unknown at the design stage and has no mathematical model. To tackle this challenge, we adopt an approximate but tractable metric called discriminant gain for classification tasks. Based on the assumption that all feature vectors follow a Gaussian mixture distribution with each Gaussian component corresponding to one class, a pair-wise discriminant gain is defined as the symmetric Kullback-Leibler (KL) divergence between an arbitrary class pair \cite{lan2022progressive}. With a larger pair-wise discriminant gain, the class pair is better differentiated in the feature space, leading to a higher achievable inference accuracy. The overall discriminant gain can be either defined as the average or the minimum of all pair-wise discriminant gains.

\subsubsection{Modeling of Sensing, Computation, and Communication}
To characterize the influence of sensing, computation, and communication on inference accuracy, it’s desirable to quantify the distortion caused during the three modules on the received feature vector at the edge server. In \cite{wen2022task}, the three modules are elaborated and modeled. To begin with, by equipping edge devices with dual-functional radar communication systems and using the frequency modulation continuous waveform (FMCW) for sensing, the echo signal can be decomposed into three parts: the desired signal directly reflected from the source target, the rich-scattered clutter signals, and the sensing noise. Typical clutter cancellation methods like singular value decomposition are then applied. Besides, since the clutter is rich scattering, it can be modeled as a Gaussian distribution according to the central limit theorem. For the computation process at devices, the sensory data is further processed using feature extractors like principal component analysis (PCA) or shallow neural networks for obtaining local feature vectors. The feature extractors are trained using offline dataset at the edge server. As for the communication process, if digital communication is applied, then feature vector quantization is further needed to make sure the transmitted bits no larger than the channel capacity. If the analog transmission is adopted, the channel fading and noise directly introduce distortion on the transmitted symbols. With the aforementioned modeling of the three modules, the mathematical model of the received feature vector at the server can be constructed and the corresponding mathematical model of inference accuracy measured by discriminant gain can be established. Accordingly, the task-oriented ISCC resource allocation scheme can be proposed for enhancing the inference accuracy under a given latency \cite{wen2022task}. 

\subsection{ISCC Schemes for Multi-device Edge Inference}
\subsubsection{Narrow-view Sensing based Multi-device ISCC schemes}
{\color{blue}Since intelligent tasks are generally complicated and require large-scale AI models, a single device has limited ability for the efficient completion of inference tasks. Specifically, it either obtains high-quality sensory data by focusing on a narrow view or acquires noise-corrupted high-dimensional sensory data by sensing a wide view.} The former case results in insufficient information and the latter causes low-quality feature vectors, both leading to inference accuracy degradation. To address this issue of narrow-view sensing, a multi-device edge inference framework is proposed in \cite{wen2022task}, where each device senses a disjoint view of a source target and extracts a local feature vector from the obtained sensory data. All local feature vectors are cascaded at an edge server to complete the downstream inference task. However, the ISCC design in such a system faces new challenges. As well as the coupling of mentioned before, the three modules of different devices compete for system resources, leading to an increased complexity. Besides, devices have heterogeneity in terms of channel gain, processing frequency, energy threshold, etc. To handle these challenges, a task-oriented ISCC scheme is proposed for the resource management among different modules of different devices \cite{wen2022task}.

\subsubsection{Wide-view Sensing based Multi-device ISCC schemes}
For the case of wide-view sensing, the split neural network (SNN) can be applied for the multi-device edge inference framework. Although a same wide view of a target is sensed by all devices, the obtained sensory data of different devices have heterogeneity due to the existence of sensing noise, interference, and angular deviation \cite{wen2022taskaircomp}. To address this issue, the SNN architecture is utilized. Specifically, an independent shallow neural network is deployed at each device for extracting a homogeneous feature vector. At the server, an aggregation layer aggregates all local vectors to generate a denoised global one, which is input into a large-scale neural network for finishing the inference task. To overcome the communication bottleneck of transmitting the local feature vectors, an accuracy-oriented broadband AirComp scheme is proposed, which targets averaging out the sensing and channel noise for maximizing the inference accuracy by simultaneously receiving all local feature vectors.

\section{Physical Layer ISCC Techniques for Supporting Edge AI}
\subsection{Vision: Motivation and challenges}
Despite the optimization of resource allocation for sensing, communication and computation processes, the competition of network resources between these processes still exists. To alleviate the competition and fully exploit the radio resource for edge AI, physical layer ISCC techniques such as beamforming are expected to be designed. As for federated learning, the ISCC signal facilitates the model training process by enabling simultaneous parameters collection at each mobile device (MD) for local model training and local results aggregation at the server for global model update, while the accuracies of collected parameters and aggregated results are guaranteed by the beamforming designs. As for edge inference, the spatial diversity is exploited by the beamforming designs to improve the target sensing accuracy for local inference while guaranteeing the accuracy of results aggregation at the server for intelligent decision making.


Designing physical layer ISCC signals faces several challenges. To support simultaneous sensing and communication, the beamformers need to be designed for balancing the radar sensing functionality evaluated by Cramer-Rao Bound or MSE and communication functionality evaluated by SNR or throughput. As for simultaneous communication and computation, the beamformers aim at equalizing the channels among MDs and suppressing noise to improve the accuracy of computed functions evaluated by MSE. For realizing simultaneous sensing, communication and computation, the beamforming design needs to account for both sensing accuracy and computation accuracy during communication. To overcome the challenges, several types of beamforming designs are introduced in the subsequent sections.

\subsection{Beamforming Design for Dual-Functional Signals}
\begin{figure*}[t]
\centering
\includegraphics[width=1\textwidth]{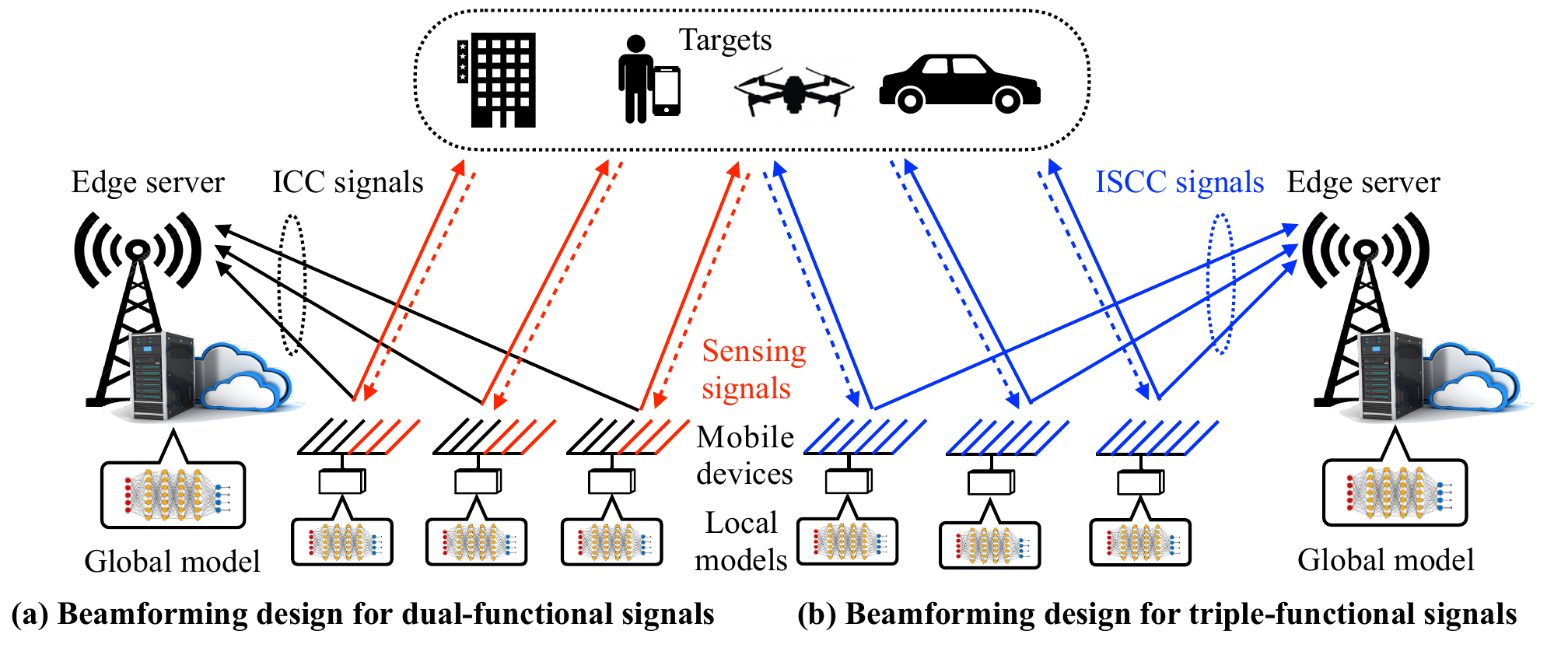}
\caption{Physical Layer ISCC for Edge AI.}
\label{FigPhysical}
\end{figure*}

As for dual-functional design, the antenna array at each MD is split for generating two signals with one for sensing and another for integrated communication and computation (ICC) as shown in Fig.~\ref{FigPhysical}(a) \cite{li2023integrated}. The information of sensing target is extracted from the statistic information of the reflected signal via maximum likelihood estimation (MLE). Meanwhile, the sensed information is computed based on the analog-wave addition of signal propagation during the data transmission from the MDs to the server. Therefore, there are two beamformers to be designed at each MD, namely radar sensing beamformer and data transmission beamformer. The sensing performance is evaluated by the MSE of the estimated target response matrix, and the ICC performance is evaluated by the MSE of the received computation results. The interference of ICC signals on radar sensing can be alleviated by the statistical methods, while the radar sensing signals will cause extra interference on ICC at the server. To improve the ICC performance, the data aggregation beamformer is deployed at the server for equalizing the received signals. The joint transmission and data aggregation beamforming design is formulated as a semidefinite programming problem for minimizing the ICC error while guaranteeing the sensing performance. The solving approach based on semidefinite relaxation is applied to obtain the desired design.

\subsection{Beamforming Design for Triple-Functional Signals}
As illustrated in Fig.~\ref{FigPhysical}(b), the ISAC and ICC are naturally combined in a triple-functional signal design \cite{li2023over}. The whole antenna array at each MD generates a single signal for realizing sensing, communication, and computation functionalities simultaneously. Compared with the dual-functional signals, there is no interference on ICC brought by radar sensing signals. {\color{blue} However, the triple-functional design brings new challenges as the corresponding transmission beamformer at each MD needs to account for both the performances of radar sensing and ICC. Pointing the beam to the sensing target will result in better sensing accuracy at the sacrifice of ICC performance. To improve the ICC performance while guaranteeing the sensing accuracy, the transmission beamformers at MDs and data aggregation beamformers are jointly designed. The resultant optimization problem can be solved by applying semi-definite relaxation and rank reduction methods.}

\begin{figure}[t]
\centering
\includegraphics[width=0.5\textwidth]{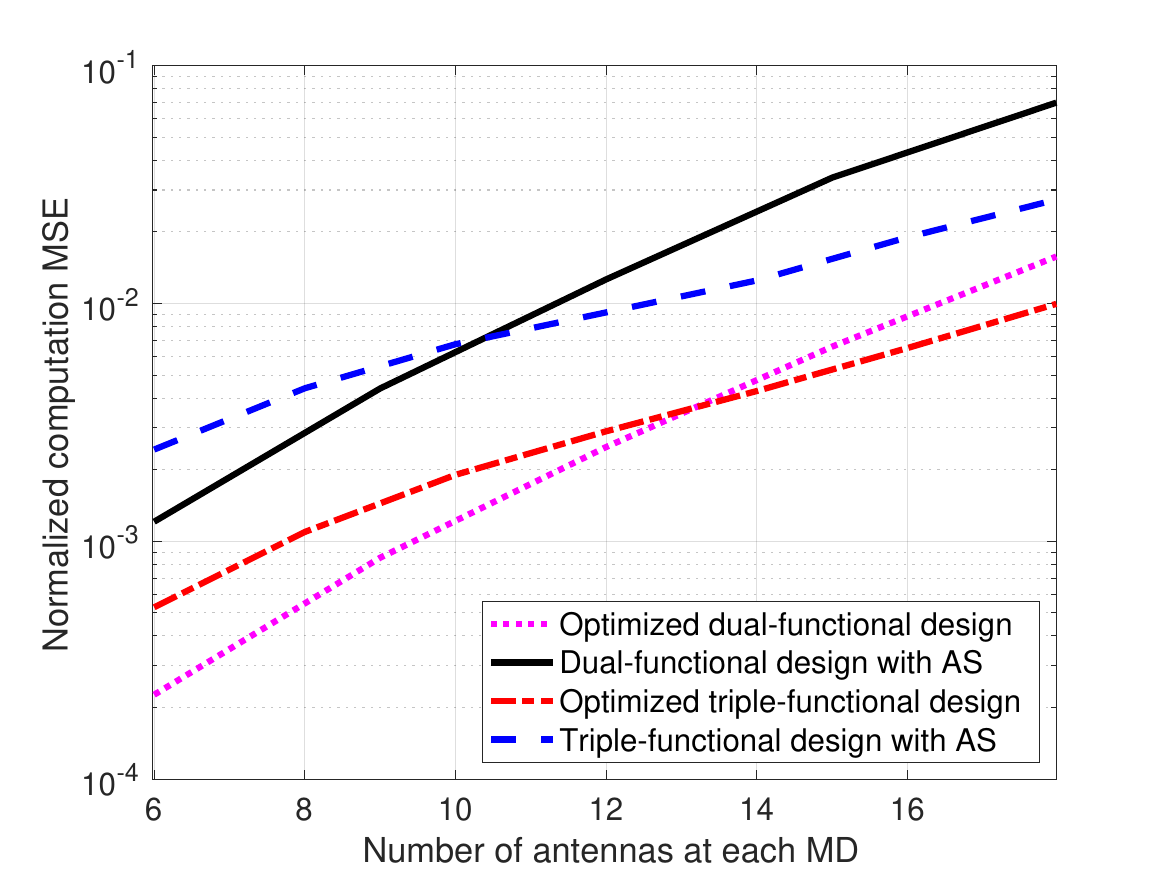}
\caption{Performance Comparison between Beamforming Designs for Dual-functional and Triple-functional Signals.}
\label{FigCompare}
\end{figure}

Fig.~\ref{FigCompare} illustrates the initial performance comparison between the beamforming designs for dual-functional and triple-functional signals. {\color{blue}There are $10$ MDs each equipped with $12$ antennas and one server with $15$ antennas. $6$ antennas are for signal transmitting and $6$ antennas are for signal receiving.} The antenna selection (AS) scheme is added as a benchmark policy, which selects the fixed amount of receive antennas with the largest channel gains. It can be observed that the ISCC performance deteriorates with the increasing number of antennas at each MD. {\color{blue}The reason is that a larger number of antennas at each MD will result in a larger dimension of target response matrix to be estimated and thus more stringent sensing constraints. Consequently, the beamformers need to sacrifice the performance of ICC to guarantee the radar sensing accuracy.} One can also observe that both the beamforming designs for dual-functional and triple-functional signals outperform the benchmark policy with AS, which illustrates the necessity of beamformer optimization. Moreover, as the number of antennas at each MD increases, the performance of the triple-functional signals becomes better than that of the dual-functional signals. The reasons are two-folded. On one hand, a larger number of antennas at each MD leads to a larger dimension of transmission beamforming matrix for supporting the triple-functionalities. On the other hand, a larger number of antennas for radar sensing at each MD will exacerbate the interference of radar sensing signals on ICC in the dual-functional design. {\color{blue}The computation MSE can be further reduced by equipping more receiving antennas at the server. Moreover, less sensors participating AirComp or less functions to be computed will result in smaller computation MSE.}

\section{Conclusion}
{\color{blue}ISCC is a novel and effective paradigm for enabling ubiquitous low-latency and reliable intelligent applications such as smart health, auto-driving, smart home, so as to empowering intelligence of everything.} Such a paradigm enjoys the advantages of resource utilization enhancement and customized goals achievement in edge AI tasks. The challenges for developing ISCC algorithms for edge AI arise from both the resource level, i.e., the highly coupled resource competence among the three modules, and the task level, i.e., the lack of unified design objectives. To this end, a set of ISCC schemes in both application layer and physical layer for several typical edge AI scenarios  are presented in this article. We hope these designs will motivate more exciting ISCC theories, methodologies, and applications for promoting the advancement of edge AI.


\bibliography{reference}
\bibliographystyle{ieeetr}

\end{document}